\documentclass[useAMS,usegraphicx]{mn2e}

\begin{document}

\title{High Accretion Rate during Class 0 Phase due to External Trigger}

\author[Kazutaka Motoyama and Tatsuo Yoshida]
       {Kazutaka
       Motoyama\thanks{E-mail: motoyama@golf.sci.ibaraki.ac.jp} and Tatsuo
       Yoshida\\
      Faculty of Science,  Ibaraki University,  Mito, Ibaraki 310-8521, Japan}

\maketitle



\begin{abstract}
Recent observations indicate that some class 0 sources have orders of
magnitude higher accretion rates than those of class I.
We investigated the conditions for the high accretion rates of
some class 0 sources by numerical calculations, modelling an external
trigger.
For no external trigger, we find that the maximum value of the accretion
rate is determined by the ratio $\alpha$ of the gravitational energy to the
thermal one within a flat inner region of the cloud core.
The accretion rate reaches $\sim 10^{-4} \rm{M_{\sun } yr^{-1}}$
if the cloud core has $ \alpha \ga 2$.
For an external trigger we find that the maximum value of the accretion
rate is proportional to the momentum given to the cloud core.
The accretion rate reaches $\ga 10^{-4} \rm{M_{\sun } yr^{-1}}$ with a
momentum of $\sim 0.1 \rm{M_{\sun} km \, s^{-1}}$ when the initial central
density of the cloud core is $\sim 10^{-18} \rm{g \, cm^{-3}}$.
A comparison between recent observational results for prestellar cores
and our no triggered collapse model indicates that the flat inner regions of
typical prestellar cores are not large enough to cause accretion rates of
$\sim 10^{-4} \rm{M_{\sun } yr^{-1}}$.
Our results show that the triggered collapse of the cloud core is more
preferable for the origin of the high accretion rates of class 0 sources
than no triggered collapse. 
\end{abstract}


   \section{INTRODUCTION}\label{sec:intro}
Pre-main-sequence stars are classified according to their spectral
features. 
Lada \& Wilking (1984) and Lada (1987) proposed a classification for
pre-main-sequence stars using the slope of their
infrared Spectral Energy Distribution (SED).
They insisted that this classification corresponds to an evolutional
sequence.
We state class I, II, and III in the order of evolution.
In addition to these classifications, Andr\'e, Ward-Thompson, \& Barsony
(1993) proposed a new category called class 0.
The sub-mm ($\lambda > 350 \rm{\mu m}$) luminosity  
of class 0 sources is $5 \times 10^{-3}$ times higher than 
total luminosities, which indicates that class
0 sources are surrounded by significantly large amounts of
circumstellar material.
Andr\'e \& Montmerle (1994) insisted that class 0 sources are
younger than those of class I.

The accretion rates of class 0 sources have been believed to be higher than
those of class I by comparing the observations and theories.  
Bontemps et al. (1996) observed the CO outflow activities of YSOs nearby
regions of star formation, and revealed that some class 0 sources have
an order of magnitude larger momentum flux than those of class I. 
Most ejection models predict that the momentum flux of the outflow is
proportional to the accretion rate (Pelletier \& Pudritz 1992; Wardle \&
K\"onigl 1993; Shu et al. 1994).  
This means that these class 0 sources have higher accretion rates than
those of class I. 
Moreover, Jayawardhana, Hartmann, \& Calvet (2001)
carried out radiative transfer calculations of infalling, dusty
envelopes surrounding embedded protostars. 
They concluded that the prototype class 0 source VLA 1623 in the $\rho$
Oph cloud must have an accretion rate
$\ga 10^{-4}\rm{M_{\sun}yr^{-1}}$ by
matching the far-infrared peak in the SED.  
This value is much higher than the typical accretion rates of $\sim 4 \times
10^{-6}\rm{M_{\sun}yr^{-1}}$ of the class I sources in Taurus-Auriga,
whose values were obtained by similar radiative transfer calculations
(Kenyon, Calvet, \& Hartmann 1993). 
Recently, Di Francesco et al. (2001) observed optically thick line
$\rm{H_2CO(3_{12}-2_{11})}$ in the class 0 source NGC1333 IRAS 4A, and
detected the inverse P-Cygni profile in the line.
They derived a
high infall velocity and a high accretion rate of $1.1 \times
10^{-4}\rm{M_{\sun}yr^{-1}}$.

One possibility is that the high accretion rates of class 0
sources are related to the initial density profiles before gravitational
collapses.
If the starting point of star formation is a singular
isothermal sphere (e.g. Shu, Adams, \& Lizano  1987), the accretion rate is
constant at $\sim c_s^3 /G$ through the entire accretion phase, where
$c_s$ and $G$ denote the sound speed and the gravitational constant,
respectively.
The time variation of the accretion rate can be caused by an
initial density profile different from that of a singular
isothermal sphere (Foster \& Chevalier
1993; Henriksen, Andr\'e \& Bontemps 1997  hereafter HAB; Whitworth \&
Ward-Thompson 2001).
HAB reproduced high accretion rates of class 0 sources by assuming that a
prestellar core has a flat inner region surrounded by a power-law envelope.
In their model, the accretion rate is high in the earlier accretion phase, and
declines later.
Their model is consistent with the picture that the accretion rates
of class 0 sources are higher than those of class I. 
  
In this paper we investigate quantitatively the conditions for the high
accretion rates of class 0 sources based on hydrodynamic calculations.
When we take into account the pressure effect,
even though the analytic HAB's model neglects it, we must assume that
the flat inner regions is more massive in order to cause high accretion rates.
Recent observations reveal both the masses and sizes of the flat inner regions
of prestellar cores (Ward-Thompson, Motte, and Andr\'e 1999).
Assuming a gas temperature of 10 K, the flat inner regions of typical
prestellar cores have thermal energies that are comparable to
the gravitational ones.
It seems that such cores do not cause high accretion rates.  

We have investigated collapses with and without an external trigger.
If star formation is triggered by an interstellar shock wave,
a high accretion rate is expected due to compression by the shock wave. 
The cloud cores interact with the shock wave under various circumstances.
For instance, triggered star formation by a distant supernova is suggested
in the Upper Sco association (Walter et al. 1994;Preibisch \& Zinnecker
1999) and the $\rho$ Oph cloud (Vrba 1977;Loren \& Wootten 1986;de Geus
1992;Preibisch \& Zinnecker 1999).
The star formation in NGC 2265 IRS in the Cone Nebula seems
to be triggered by stellar wind from a B2 star (Thompson et al. 1998). 
The triggered formation scenario is also proposed for the
origin of the solar system based on studies of meteorites
(Cameron \& Truran 1977).
The presence of short-lived isotopes in meteorites, which should be
provided by stellar ejecta from a supernova, or an asymptotic giant
branch star, leads to this scenario.

Many numerical simulations have been carried out to investigate
the triggered collapse scenario (Boss 1995; Foster \& Boss 1996; Vanhala
\& Cameron 1998). 
Hennebelle et al. (2002) investigated the evolutions of a Bonnor-Ebert
sphere which is subjected to an increase in the external pressure.
They demonstrated that the features of the density and
velocity field of prestellar cores and protostars are well reproduced,
and that the accretion rate is high in the earlier accretion phase.
In this paper, we examine the possibility that an external trigger
causes a high accretion rate of $\sim 10^{-4}\rm{M_{\sun}yr^{-1}}$.
We investigate the conditions for the high accretion rates of class 0
sources, pursuing numerical calculations of the collapse of the cloud
core triggered impulsively by an external shock wave.

In Section \ref{sec:models}, we describe our models and numerical method.
In Section \ref{sec:result}, we present the numerical results.
In Section \ref{sec:discussion}, we summarize the main results and discuss
them.

   \section{Models and Numerical Method}\label{sec:models}
   \subsection{Basic equations and models}\label{ssec:basiceq}
The fluid equations in spherical coordinates are
\begin{equation}
\frac{d \rho}{d t}+ 4 \pi \rho^2   
\frac{\partial}{\partial m}(r^2 v)  =0,
\label{spmasscons2}
\end{equation}

\begin{equation}
\frac{d v}{d t} = - 4 \pi r^2  
\frac{\partial P}{\partial m}-\frac{Gm(r)}{r^2},
\label{speqofmotion2}
\end{equation}
where $\rho$ is the density, $v$ the radial velocity, $P$ the
pressure, and $m$ the mass within radius $r$,
\begin{equation}
m(r) =\int^r_0 4 \pi {r'}^2 \rho (r') dr'.
\label{mass}
\end{equation}
Although the density of the cloud core increases drastically in the
process of star formation, radiative cooling is effective except for the
high-density region in the centre (Larson 1969). 
We assume that the medium is an isothermal gas, 

\begin{equation}
P=c_s^2 \rho.
\label{isother}
\end{equation}
The gas temperature T is set at 10 K ($c_s = 215 \rm{m \, s^{-1}}$),
which is the typical temperature for globules and cloud cores
(e.g. Clemens, Yun, and Heyer 1991) in our calculations.
The mean molecular weight and the ratio of the specific heats are set at 2.5
(assuming cosmic abundance) and 7/5, respectively.

We adopt a Plummer-like density profile as an initial condition (Whitworth \&
Ward-Thompson 2001).
Recent high angular resolution observations of the precollapse
cloud support this Plummer-like density profile.
Ward-Thompson et al. (1994) demonstrated that prestellar cores have flat
inner regions surrounded by power-law envelopes based on the observation of
starless $\rm{NH_3}$ cores.
We assume the following initial density profile:

\begin{equation}
\rho_0 (r) =\rho_{flat} {\left[  \frac{R_{flat}}
{ {\left(  R_{flat}^2 + r^2  \right)}^{\frac{1}{2}}} \right]}^{\eta}.  
\label{speqofmotion2}
\end{equation}
At a radius smaller than $R_{flat}$, the density is nearly constant,
$\rho_{flat}$. At a radius larger than $R_{flat}$, the density is
proportional to $r^{-\eta}$.
We fix the parameter $\eta = 2$ in our models presented in section 3.

   \subsection{Numerical methods}\label{ssec:method} 

\begin{figure}
\includegraphics[height=7cm]{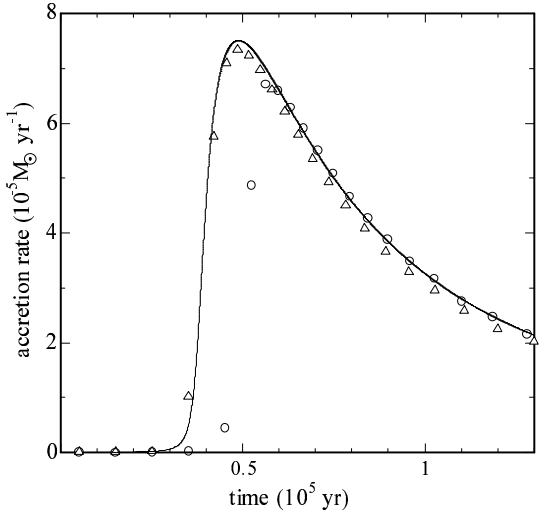}
\caption{Accretion rate as a function of time.
The solid line denotes the analytic solution given by Whitworth \&
 Ward-Thompson (2001).
The circles and the triangles denote the case of $T=$10 K and $T=$ 0.5 K, respectively. }
\label{codetest}
\end{figure}

We solved numerically equations (\ref{spmasscons2})-(\ref{isother})
using a finite difference scheme.
These equations were integrated by Godunov's method.
This Lagrangian hydrodynamical code has a second order accuracy in both
space and time.
We set the inner ($r=0$) boundary conditions such that
$P$ and $\rho$ have a zero gradient, and $v=0$, and the outer boundary
conditions that $P$, $\rho$, and $v$ have a zero gradient.
We used the non-uniform grids, $\Delta m_{i+1} = (1+\epsilon)\Delta
m_i$, where $\Delta m_{i}$ is the  mass inside the $i$th grid from the centre.
The value of $\epsilon$ was set at $3 \times 10^{-3}$ in order to obtain
a high resolution at the centre.
We used 6000 grids in our calculations.
For the cloud core itself of the mass 3 $M_{\sun}$, 2737 grids were allocated.
Outside of the cloud core, the density is equal to that of core edge,
and the gravitational force was excluded from our calculation. 

We used the sink-cell method adopted by Boss \& Black (1982) to
calculate the later accretion phase.
In many numerical simulations which treated the collapse of the isothermal
sphere, the density and the infall velocity of the central region increase
drastically during a runaway collapse phase. 
It is difficult to continue calculations to a later stage, because the time
step that satisfies the CFL condition is too small. 
When the central density reaches a reference density value, $\rho_{sink}$,
cells within radius $r_{sink}$ are treated as sink-cells:
the accreted gas into sink-cells is treated as point like mass at
the centre.
Since the infall velocity near the sink-cell exceeds the sound
velocity, the effect of the sink-cell does not travel to the outer gas. 

As a test for the numerical code, we compared our result with the analytic
solution of Whitworth \& Ward-Thompson (2001).
We evaluated the accretion rate at the radius $r_{acc}=$ 300 AU.
Fig. \ref{codetest} shows comparison between the analytic solution and our
results.
Each parameter was set as follows: $\rho_{flat} = 3.0 \times 10^{-18} \,
\rm{g \, cm^{-3}}$, $R_{flat}=5350$ AU, and $\eta =$ 4.
The gas temperatures are set at 10 K and 0.5 K.
The result in the case of $T=$ 0.5 K agrees well with the analytic solution.
On the other hand, in the case of $T=$ 10 K,
the maximum value of accretion rate is less than that of the analytic solution 
because of a pressure effect.

   \section{RESULTS}\label{sec:result}
We studied two different models in order
to seek for the conditions for a high accretion rate of $\sim 10^{-4}
{\rm M_{\sun } yr^{-1}}$.
In Section \ref{ssec:spontaneous} and Section \ref{ssec:triggered}, we
consider the collapse of cloud core without and with an external trigger,
respectively.
We call the former the no triggered collapse model, and the latter triggered
collapse model in this paper.
       \subsection{No triggered collapse model}\label{ssec:spontaneous}

\begin{table}
\caption{Model parameters and the maximum value of accretion rate
 $\dot{M}_{max}$ for the no triggered collapse model.} 
\begin{center}
$$
\begin{array}{cccccc}

   &  \rho_{flat} &  R_{flat} & M_{flat}  & \alpha & \dot{M}_{max} \\ 

model   &  (\rm{g \, cm^{-3}}) &  (\rm{AU}) & (\rm{M_{\sun}}) &  &
(\rm{M_{\sun} yr^{-1}}) \\

\hline\hline

A1   & 1.0 \times 10^{-18} &  3400 & 0.18 & 0.42  & 4.27\times10^{-5} \\
A2   & 1.0 \times 10^{-16} &   710 & 0.18 & 1.8  & 1.82\times10^{-4} \\
B1   & 1.0 \times 10^{-18} &  4300 & 0.36 & 0.67  & 5.45\times10^{-5} \\
B2   & 1.0 \times 10^{-16} &   930 & 0.36 & 3.1  & 3.88\times10^{-4} \\
C    & 2.0 \times 10^{-18} &  4200 & 0.67 & 1.3  & 9.77\times10^{-5} \\
D1   & 1.0 \times 10^{-18} &  8000 & 2.3  & 2.3  & 1.59\times10^{-4} \\
D2   & 1.0 \times 10^{-16} &  1720 & 2.3  & 11   & 1.56\times10^{-3} \\

\end{array}
$$
\end{center}

\label{inicon}
\end{table}

\begin{figure}
\includegraphics[width=8cm]{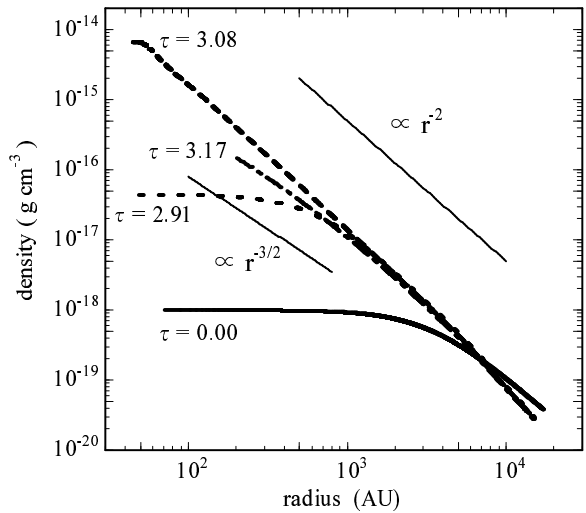}
\caption{Evolution of the density profile in model A1.
The number attached to each line denotes the time normalized by the
 initial free fall time at the centre.}
\label{nosdensity}
\end{figure}

\begin{figure}
\includegraphics[width=8cm]{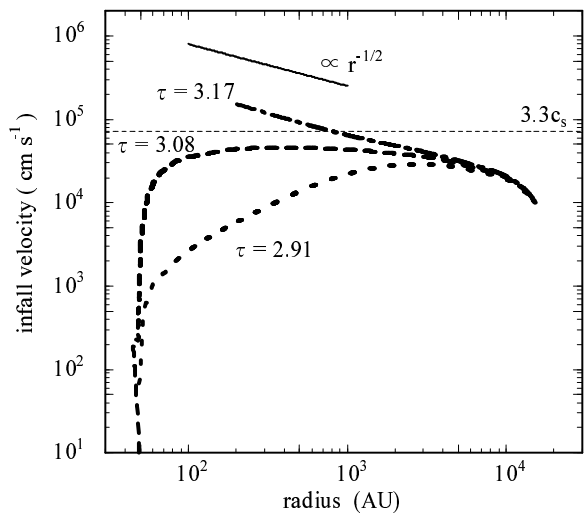}
\caption{Evolution of the velocity profile in model A1. 
The number attached to each line is the same as in Fig. \ref{nosdensity}.}
\label{nosvelocity}
\end{figure}

Here, we present the numerical results of the no triggered collapse model.
We summarize the model parameters and the maximum value of
accretion rate in Table \ref{inicon}.
Fig. \ref{nosdensity} and Fig. \ref{nosvelocity} show the evolutions of the
density and the velocity profiles in model A1.
Here, we use the time $\tau = t \, / \, t_{ff}$, which is normalized
by the initial free fall time, $t_{ff}=\left( \frac{3 \pi}{32 G
\rho_{flat}} \right)^{1/2}$, at the centre.
In the runaway collapse phase ($\tau < 3.09$), the flat inner region
shrinks and its density increases as the collapse proceeds. 
The density is proportional to $r^{-2}$ in the outer envelope. 
The infall velocity is nearly uniform at $\sim 3.3 \, c_s$ in the outer
region.
After core formation ($\tau > 3.09$), the density and the infall velocity
are proportional to $r^{-3/2}$ and $r^{-1/2}$ in the central region,
respectively.

\begin{figure}
\includegraphics[height=7cm]{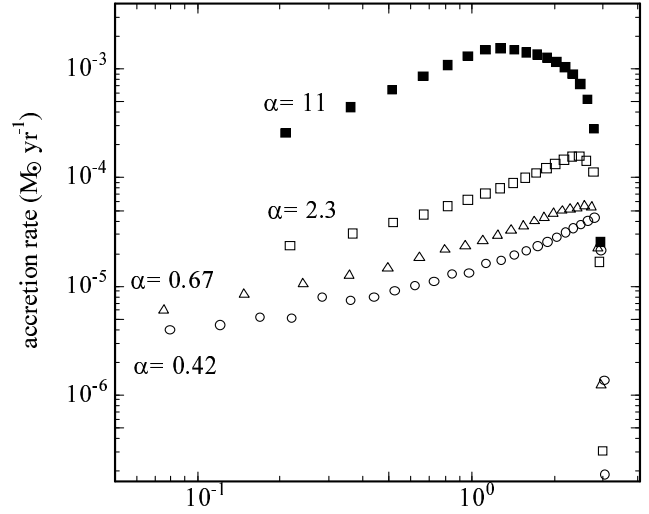}
\caption{Accretion rate as a function of the envelope mass. The
 open circles, open triangles, open squares, and filled squares denote
 model A1, B1, D1, and D2, respectively. The number attached to each
 plot denotes the ratio $\alpha$ of the gravitational energy to the thermal one
 within the inner flat region.} 
\label{nosaccrate}
\end{figure}

Fig. \ref{nosaccrate} shows the accretion rates of each model as
a function of the envelope mass, which is used as an evolutionary indicator. 
The envelope mass denotes the mass that remains outside of radius $r_{acc}$. 
In all models, the accretion rates increase rapidly during the earlier
collapse phases.
After the accretion rate reaches the maximum value, $\dot{M}_{max}$, it
gradually declines.
The maximum value of accretion rate increases as the ratio $\alpha$ of
the gravitational energy to the thermal one within the inner flat region
is large.
Fig. \ref{nosaccrate} indicates that the value of $\alpha$ is larger
than 2 in order to cause an accretion rate of $\sim 10^{-4} {\rm
M_{\sun } yr^{-1}}$.
We compared our results with the accretion rates of the
pressure-free models.
For models with $\alpha \ga 2$, the maximum values of the accretion
rates can be fitted well with pressure-free solutions.

We can estimate the accretion rate for a larger $\alpha$, neglecting pressure
effect.
The inner flat region with mass $M_{flat}$ collapses within 

\begin{equation}
 \Delta t = \overline{t_{ff}}(R_{flat}) - t_{ff} 
= \left( 1 - \frac{\sqrt{3(4-\pi)}}{2} \right) \overline{t_{ff}}(R_{flat}),
\end{equation}
where $\overline{t_{ff}}(R_{flat})$ represents the free fall time of the
mean density, $\overline{\rho}(R_{flat})=[4 \pi \rho_{flat}
R_{flat}^3 (1-\frac{\pi}{4})]/(\frac{4}{3} \, \pi R_{flat}^3)$,
within the radius
$R_{flat}$.
The accretion rate is given by 

\begin{eqnarray}
 \dot{M} & \sim & \frac{M_{flat}}{\Delta t}\\
         & = & \frac{2}{\pi \left( 1 - \frac{\sqrt{3(4-\pi)}}{2} \right)} 
               \left( \frac{2GM_{flat}^3}{R_{flat}^3} \right)^{1/2}. 
                    \label{dm/dt}  
\end{eqnarray}
If we consider the inner flat region as approximately an uniform sphere, the
ratio $\alpha$ is estimated to be  
\begin{eqnarray}
 \alpha & \sim & 2GM_{flat}/ (5c_s^2 R_{flat}) \nonumber \\
        & \sim & 0.42 
              \left( \frac{M_{flat}}{0.18 \, M_{\sun}} \right)
              \left( \frac{R_{flat}}{3400 \, \rm{AU}} \right)^{-1} 
              \left( \frac{T}{10 \, \rm{K}} \right)^{-1}.
\label{alpha}
\end{eqnarray}
Substituting this relation into equation (\ref{dm/dt}) gives
\begin{equation}
 \dot{M} \sim  4.24 \times 10^{-5} \alpha^{3/2}   \,
               \left( \frac{T}{10 \, \rm{K}} \right)^{3/2}
                  \,\,\,   \rm{M_{\sun} \, yr^{-1}}.
\label{dotm}
\end{equation}
This estimation neglecting pressure effect is appropriate for the cases
with $\alpha \ga 2$.
Then, the accretion rates are essentially determined by $R_{flat}$ and
$M_{flat}$, although equation (\ref{dotm}) seems to depend on $T$.
A similar estimation is also found in Ogino, Tomisaka, \& Nakamura (1999).

       \subsection{Triggered collapse model}\label{ssec:triggered}

\begin{table}
\caption{Model parameters, the arrival times at the centre of shock
waves $t_s$, and the maximum values of the accretion
rate for triggered collapse models.
The value of the initial free fall time at the centre $t_{ff}$ is
$6.65 \times 10^4 {\rm yr}$ in the case of $\rho_{flat}=1.0 \times
 10^{-18} {\rm g \, cm^{-3}}$.}
\begin{center}

$$
\begin{array}{ccccc}

    & v_{ini}  & M_{ini}  & t_s   &  \dot{M}_{max} \\

model    & ({\rm km \, s^{-1}}) & (10^{-2} \times {\rm M_{\sun}}) &
 (\times t_{ff})    & ({\rm M_{\sun} yr^{-1}})

\\ \hline\hline

D3   &  3     & 0.9  & 3.16  & 6.73\times10^{-5} \\

E5   &  5     & 1.8  & 2.93  & 1.09\times10^{-4} \\

E10  &  10    & 1.8  & 2.61  & 1.10\times10^{-4} \\

E15  &  15    & 1.8  & 2.35  & 1.23\times10^{-4} \\

E20  &  20    & 1.8  & 2.15  & 1.34\times10^{-4} \\

F5  &  5      & 3.6  & 2.63  & 1.11\times10^{-4} \\

F10  &  10    & 3.6  & 2.16  & 1.40\times10^{-4} \\

F15  &  15    & 3.6  & 1.84  & 1.99\times10^{-4} \\

F20  &  20    & 3.6  & 1.60  & 2.65\times10^{-4} \\

G10  &  10    & 4.5  & 1.99  & 1.63\times10^{-4} \\

G15  &  15    & 4.5  & 1.68  & 2.47\times10^{-4} \\

\end{array}
$$
\end{center}

\label{results}

\end{table}
Here, we present numerical results of the triggered collapse model. 
We give some grids outside of the cloud core the velocity to a centre,
$v_{ini}$.
The total amount of mass in these grids is $M_{ini}$.
A similar method is used in Boss (1995).
We summarize the model parameters, the arrival times at the centre of shock
waves $t_s$, and the maximum values of the accretion rates in Table
\ref{results}.
We adopt the density distribution of model A1 as an initial
condition, of which $\alpha$ is the smallest in no triggered collapse models.

\begin{figure}
\includegraphics[height=7cm]{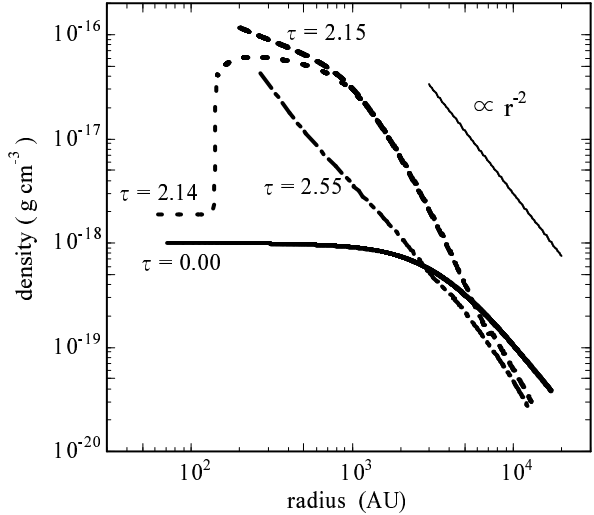}
\caption{Evolution of the density profile in model E20.
The number attached to each line is the same as in Fig. \ref{nosdensity}.} 
\label{sdensity}
\end{figure}

\begin{figure}
\includegraphics[height=7cm]{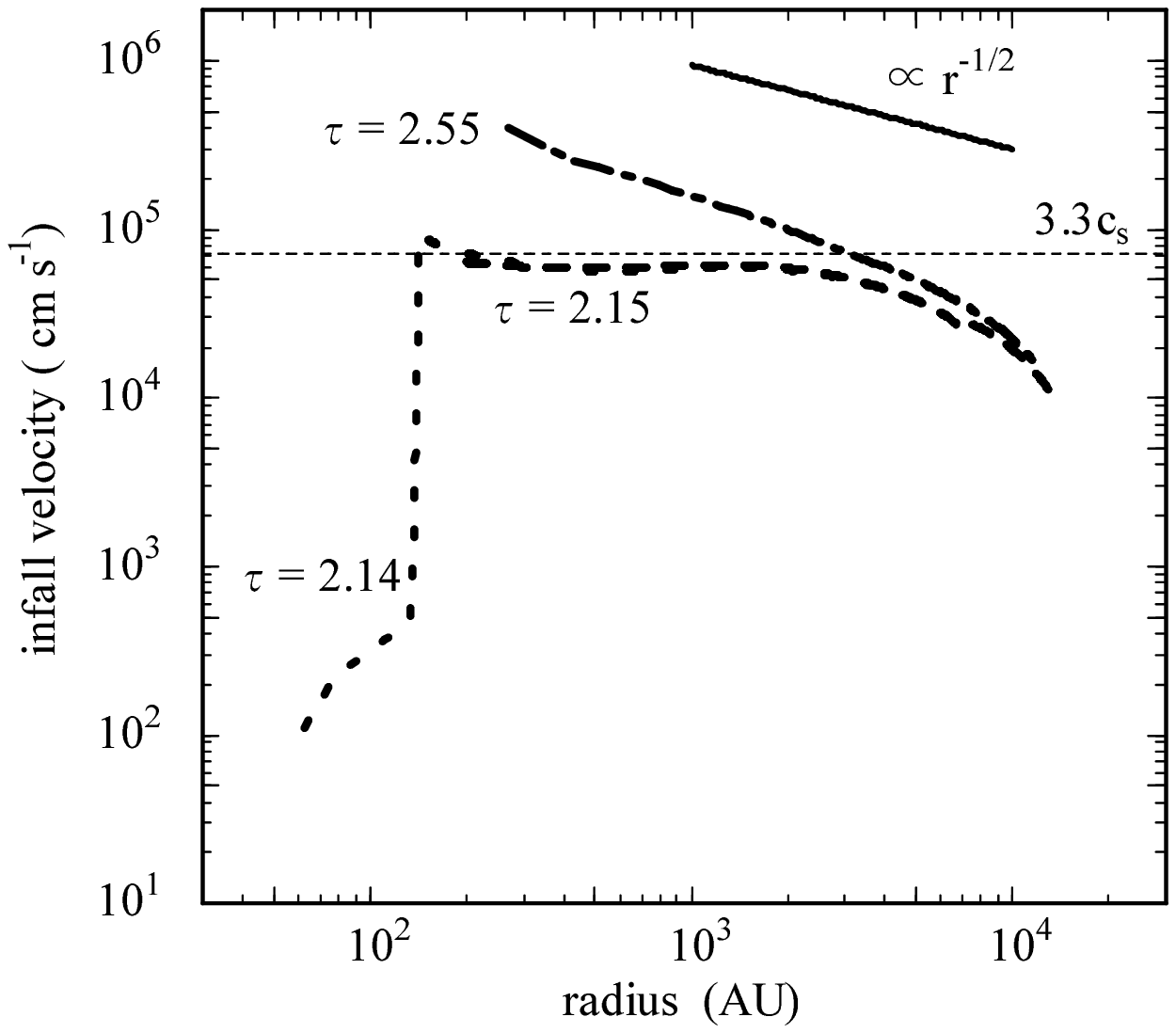}
\caption{Evolution of the velocity profile in model E20. 
The number attached to each line is the same as in Fig. \ref{nosdensity}.}
\label{svelocity}
\end{figure}

Fig. \ref{sdensity} and Fig. \ref{svelocity} show the evolutions of the
density and the velocity profiles in model E20, respectively. 
While the shock front propagates to the centre ($\tau < 2.14$), the shock wave
compresses the envelope gas.
As a result, the cloud core has a centrally concentrated configuration
relative to the initial state.
The density of this region is approximately uniform and one order higher
than the density of the preshock region, where the shock wave has not
yet arrived, and the density  profile of the outer region is steeper than
$\propto r^{-2}$.
The propagating speed of the shock front is nearly constant during
compression. 
The velocity of the preshock region is nearly at rest, and the velocity of
the postshock region is uniform.
Compared with no triggered model, the envelopes are denser and have steeper
outer-regions during the earlier collapse phases, which correspond to
the class 0 phases.

\begin{figure}
\includegraphics[height=7cm]{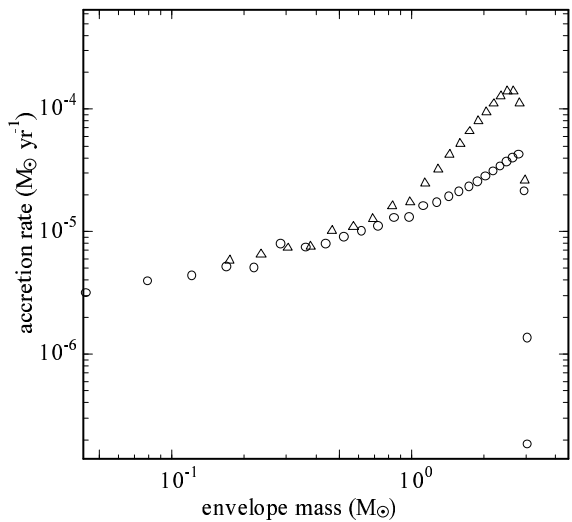}
\caption{Accretion rates as a function of the envelope mass. The
 open circles and the open triangles denote models A1 and E20,
 respectively.}
\label{sacc}
\end{figure}

Fig. \ref{sacc} represents the accretion rate in model E20 compared
with model A1.
In model E20, the accretion rate takes the maximum value ($\tau =
2.24$) just after the collapse of the flat inner region. 
After that, the accretion rate declines rapidly. 
In the later phase there is no large difference between the accretion rates
of model A1 and model E20.
Other models have the same time evolution as model E20.
 
\begin{figure}
 \includegraphics[height=7cm]{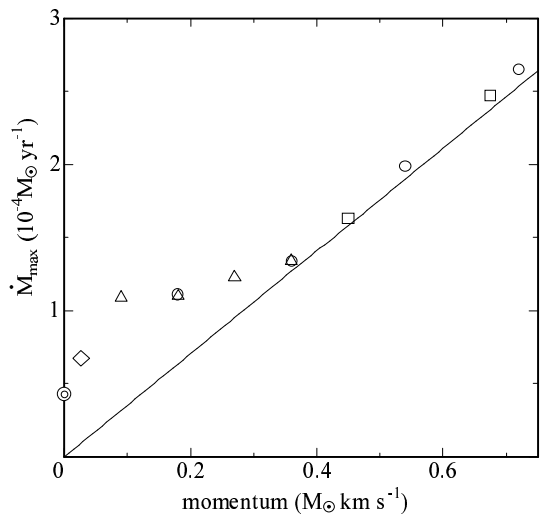}
\caption{Maximum values of the accretion rates $\dot{M}_{max}$ versus
the momenta given to cloud cores, $M_{ini} v_{ini}$.
The double open circle denotes model A1. 
The open diamond denotes model D3.
The open triangles denote models E5, E10, E15, and E20.
The open circles denote models F5, F10, F15, and F20.
The open squares denote models G10 and G15.
The solid line denotes the relation of equation (\ref{dotm-mom}). }
\label{moment}
\end{figure}

Our results suggest that the maximum values of the accretion rates are
proportional to momenta given to the cloud cores. 
Fig. \ref{moment} plots the maximum values of the accretion rates versus the 
momenta given to the cloud cores, $M_{ini} v_{ini}$. 
The models with the same momenta show similar values of
$\dot{M}_{max}$and $t_s$ : i.e., E10 and F5, E20 and F10.
The maximum value of accretion rate is determined by the momentum given
to the cloud core.
A momentum $\ga 0.1 {\rm M_{\sun} km \, s^{-1}}$ causes an accretion rate of 
$\ga 10^{-4} \rm{M_{\sun} yr^{-1}}$.

The maximum value of the accretion rate can be estimated as follows.
From equation (\ref{alpha}), $\alpha$ of the centrally concentrated flat region
compressed by external shock wave is
\begin{equation}
 \alpha   \sim  \frac {2G}{5c_s^2} \frac {M_{com}}{R_{com}}.
\label{alpha-trigger}
\end{equation}
Here, $M_{com}$ and $R_{com}$ denote the mass and the radius of this
compressed region, respectively.
If the arrival time at the centre of the shock wave $t_{s}$ is shorter than
the time $t_{nt}$, when the central density reaches $\rho_{sink}$, for
the no triggered case (i.e. model A1), we can assume that the density and the
velocity of the preshock region remain unchanged from the initial state.
Using the relation of the compression ratio for isothermal shock, we
obtain $\rho_{com}/\rho_{flat} \sim (v_{com}/c_s)^2$, where
$\rho_{com}$ and $v_{com}$ denote the density and the velocity of the
compressed region, respectively.
Thus, equation (\ref{alpha-trigger}) is rewritten as 
\begin{equation}
     \alpha  \sim  \frac{2 G}{5 c_s^2} \left( \frac{4 \pi \rho_{flat}}
                  {3 c_s^2 }  \right)^{1/3} (M_{com}v_{com})^{2/3}.
\label{alpha-comp}
\end{equation}
The momentum within the compressed flat region is conserved if the impulse
$I_{pre}$ given from the preshock region is negligible.
This impulse 
\begin{equation}
 I_{pre} = \int_0^{t_{s}} \! \! 4 \pi r_s^2 c_s^2 \rho_0(r_s)  dt 
\end{equation}
is negligible compared with $M_{ini} v_{ini}$, assuming a constant
velocity of the shock front (i.e. $r_s=R_c (1-t/t_s)$), where $r_s$ and
$R_c=1.7 \times 10^4 \, \rm{AU}$ denotes the distance from the centre to
the shock front and the radius of the cloud core, respectively.
We can replace $M_{com}v_{com}$ with $M_{ini} v_{ini}$ in equation
(\ref{alpha-comp}).
Substituting this relation into equation (\ref{dotm}) gives
\begin{eqnarray}
 \dot{M} \sim  3.52 \times 10^{-4} 
            \left( \frac{M_{ini} v_{ini}}{1 \, M_{\sun} {\rm \, k m \, s^{-1}}}
                  \right)
            \left( \frac{\rho_{flat}}{10^{-18} \, {\rm \, g \, cm^{-3}}}\right)
                       ^{1/2} \nonumber  \\
             \times \left( \frac{T}{10 \, {\rm K}} \right)^{-1/2}
               \,\,\,  {\rm M_{\sun} \, yr^{-1}}.
\label{dotm-mom}
\end{eqnarray}
As shown Fig. \ref{moment}, our results are classified into three
categories according to the comparison between the arrival time at the
centre of the shock wave $t_s$ and the required time for the central
density to reach $\rho_{sink}$, $t_{nt}=3.11 \, t_{ff}$.
In the case of $t_s \ll t_{nt}$ ($M_{ini} v_{ini} \ga 0.3
{\rm M_{\sun } km \, s^{-1}}$), our numerical results 
agrees well with the relation of equation (\ref{dotm-mom}).
In the case of $t_s \sim t_{nt}$ ($0.05
{\rm M_{\sun } km \, s^{-1}}  \la M_{ini} v_{ini} \la 0.3
{\rm M_{\sun } km \, s^{-1}}$), the maximum values of the accretion rates
are higher than the relation of equation (\ref{dotm-mom}), because there
are effects of both compression by the shock wave and contraction by
self-gravity.
In the case of $t_s > t_{nt}$ ($M_{ini} v_{ini} \la 0.05
{\rm M_{\sun } km \, s^{-1}}$), the effect of the compression by the
shock wave is small.

   \section{SUMMARY and DISCUSSION}
     \label{sec:discussion}
We summarize our main results as follows:
\begin{enumerate}
 \item The collapse of cloud core with $\alpha \ga 2$ can be treated
       as pressure-free, and the maximum value of the accretion rate is
       proportional to $\alpha^{3/2}$ in the no triggered collapse, where
       $\alpha$ is the ratio of the gravitational energy to the thermal
       one within the flat inner region of the cloud core.
       A flat inner region with $\alpha \ga 2$ causes an accretion rate of 
       $\ga 10^{-4} \rm{M_{\sun } yr^{-1}}$. 
 \item For triggered collapse, the cloud core has a centrally concentrated
       configuration relative to the initial state as the result of
       compression by the shock wave.
       Just after the shock front reaches the centre, when the cloud
       core has compressed the flat inner region and outer steep envelope,
       the accretion rate attains the maximum value.
 \item The maximum value of the accretion rate is proportional to the momentum
       given to the cloud core in the triggered collapse.
       The momentum $\ga 0.1 \rm{M_{\sun} km \, s^{-1}}$ causes
       an accretion rate of $\ga 10^{-4} \, \rm{M_{\sun } yr^{-1}}$ when
       the initial central density of the cloud core, $\rho_{flat}$, is
       $\sim 10^{-18} \rm{g \, cm^{-3}}$.
\end{enumerate}

Hennebelle et al. (2003) investigated the evolutions of a Bonnor-Ebert
sphere which is subjected to an increase in the external pressure with
Smoothed Particle Hydrodynamics simulations.
Their simulations indicate that the prestellar core has a flat inner region and
a steep outer envelope, and approximately a uniform infall velocity profile.
Their results for the fast compression cases, $\phi < 1$, are similar to
ours concerning the density and the velocity  profiles, where their
parameter $\phi$ is defined as the ratio of the time scale on which the
external pressure doubles to the initial sound crossing time. 
These features are consistent with observations of prestellar cores.
However, in their calculations the maximum value of the accretion rate
reaches only $\sim 3 \times 10^{-5} \rm{M_{\sun } yr^{-1}}$.
This low accretion rate is due to their low initial central density,
$\rho_{flat} \sim 6 \times 10^{-20} \rm{g \, cm^{-3}}$, which is two
orders lower than that of our model, $\rho_{flat} =  10^{-18}
\rm{g \, cm^{-3}}$.

We discuss the masses and the sizes of the flat inner regions of
prestellar cores.
Although in the $\rho$ Oph cloud there is a high-accretion class 0 source, i.e.
VLA 1623, a comparison between the observational results and our no
triggered collapse model indicates that the flat inner regions of
prestellar cores are not large enough to cause accretion rates of
$\ga 10^{-4} \rm{M_{\sun } yr^{-1}}$.
The mass and the radius of the flat inner region of
the prestellar core L1689B in the $\rho$ Oph complex
are estimated to be $\sim$ 0.33 $M_{\sun}$ and $\sim 4000 \rm{AU}$
(Andr\'e, Ward-Thompson, \& Motte 1996).
The parameters of this prestellar core nearly coincide with those of
model B1, in which the accretion rate and $\alpha$ are $ 5.45
\times10^{-5} \rm{M_{\sun} yr^{-1}}$ and 0.67, respectively.
Moreover, Ward-Thompson et al. (1999) recently observed eight isolated
prestellar cores.
They estimated the typical mass and the radius of the flat inner
regions to be  $\sim$ 0.7 $M_{\sun}$ and $\sim 4000 \rm{AU}$.
The parameters of this prestellar core nearly coincide with those of
model C, in which the accretion rate and $\alpha$ are $ 9.77
\times10^{-5} \rm{M_{\sun} yr^{-1}}$ and 1.3, respectively.
These results indicate the difficulty of evolution from the typical
prestellar core into the observed high-accretion class 0 source without
an external trigger.
 
In regions where stars form in cluster, some observational results
indicate the triggered collapse scenario.
Tachihara et al. (2002) found that clouds with cluster formation have
head-tail structures, studying statistically 179 cloud cores in nearby
star-forming regions.
Such structures indicate interactions with the external shock wave. 
Moreover, the class 0 envelopes in Perseus are 3 to 12
times denser than singular isothermal sphere at $T=$ 10 K (Motte and
Andr\'e 2001).
The high density of class 0 sources can be interpreted as being the
results of compression by an external shock wave.
These facts suggest that star formations are caused by external triggers
in these regions.

Momentum plays an important role in triggered star formation.
Foster \& Boss (1996) investigated the interaction of a plane wave and
the cloud core by two-dimensional calculations.
They showed that a fast adiabatic shock wave is liable to
disrupt the cloud core, and that a slow isothermal shock wave causes a
collapse of the cloud core.
They also showed that the momentum given to the cloud core is more essential
than the kinetic energy in predicting whether
induced collapse occurs or not.

Next, we discuss the magnitude of the momentum in triggered collapse.
Foster \& Boss (1996) estimated the momentum given to the cloud core by
an explosion with an initial kinetic energy of $E = 2 \times 10^{50}
\rm{erg}$.
They assumed that the evolution of a remnant switches from the Sedov-Taylor
expansion phase to an isothermal expansion phase at a shell temperature of
$10^6$ K.
They adopted a mean molecular weight of $\mu =0.61$.
This switch occurs at a radius of $\sim 14 \, \rm{pc} 
(E / 2 \times 10^{50} \, \rm{erg} )^{1/3} ( n / 1
\, \rm{cm^{-3}} )^{-1/3}$.
The momentum impacting on the cloud core at a distance $R=29$ pc from a
supernova is estimated to be $\sim 0.1 \, \rm{M_{\sun} km \, s^{-1}} \, (
E / 2 \times 10^{50} \, \rm{erg} ) ( R_c / 1.7 \times 10^4
\, \rm{AU} )^2 
( R / 29 \, \rm{pc} )^{-2}$, where $R_c$ is the radius of
the cloud core in our triggered collapse model.
Therefore, a supernova located up to 29 pc away from cloud cores would
cause a high accretion rate of $\ga 10^{-4} \, \rm{M_{\sun } yr^{-1}}$.
Although the cloud core is impacted by a planar shock in this situation,
our calculation is spherically symmetric.
If the pressure of the post-shock region is sufficiently comparable to that
of the cloud core, and the shock wave does not significantly compress before
he shock front passes the core (i.e. $t_s > 2 R_c / v_{ini}$), the cloud core
would be spherically compressed.
Otherwise, the cloud core would be compressed from one side (Lim,
Hartquist, \& Williams 2001).
The accretion rate would have a somewhat lower value than in the case of
spherically symmetry.
In the $\rho$ Oph cloud scenarios of shock triggered star formation
suggested by many authors (Vrba 1977; Loren \& Wootten 1986; de Geus
1992; Preibisch \& Zinnecker 1999), 
an expanding shell of H\,{\scriptsize I} gas created by a supernova in the
Sco OB2 association induces star formation in the $\rho$ Oph cloud at a
distance of about 20 pc from the Upper-Scorpius subgroup of the Sco OB2.  

We have investigated the conditions for the high accretion rates of
class 0 sources. 
Our results indicate that it is difficult to cause a high
accretion rate of $\ga 10^{-4} \rm{M_{\sun } yr^{-1}}$ with no
triggered collapses of typical prestellar cores.
For no triggered collapse, the flat inner region of prestellar core must be
$\alpha \ga 2$ in order to cause this
high accretion rate.
Otherwise, an external trigger is necessary for a high accretion rate.
Future high-resolution observations (e.g. ALMA) will reveal the detailed
density and velocity structures of prestellar and protostellar
cores. Observations with high velocity resolution give accurate
accretion rates of protostars from their line profiles. If
we know the detailed structures of the envelopes and accretion rates of class
0 sources simultaneously, it would reveal whether an external trigger
affect the accretion rate.

\section*{ACKNOWLEDGEMENTS}
We thank an anonymous referee for many helpful comments concerning our
manuscript.


\end{document}